\documentclass[a4paper,fleqn,usenatbib]{mnras}

\usepackage{newtxtext,newtxmath}

\usepackage[T1]{fontenc}
\usepackage{ae,aecompl}

\usepackage{graphicx}	
\usepackage{amsmath}	
\usepackage{amssymb}	

\defcitealias{SLC2015}{SLC15}

\title[Nearby strong lensing elliptical galaxies]{Improved mass constraints for two nearby strong-lensing elliptical galaxies from Hubble Space Telescope Imaging}

\author[Collier W. et al.]{
William P. Collier,\thanks{E-mail: william.p.collier@durham.ac.uk}
Russell J. Smith,
John R. Lucey
\\
Centre for Extragalactic Astronomy, Departments of Physics, University of Durham, Durham DH1 3LE, UK\\
}

\date{MNRAS Accepted 1 September 2017. Received 1 September 2017 ; in original form 2 August 2017}

\pubyear{2017}

\begin{document}
\label{firstpage}
\pagerange{\pageref{firstpage}--\pageref{lastpage}}
\maketitle

\begin{abstract}

We analyse newly obtained \textit{Hubble Space Telescope} (HST) imaging for two nearby strong lensing elliptical galaxies, SNL-1 (z = 0.03) and SNL-2 (z = 0.05), in order to improve the lensing mass constraints. The imaging reveals previously unseen structure in both the lens galaxies and lensed images. For SNL-1 which has a well resolved source, we break the mass-vs-shear degeneracy using the relative magnification information, and measure a lensing mass of 9.49\,$\pm$\,0.15 $\times$\,10$^{10}$ M$_{\odot}$, a 7 per cent increase on the previous estimate. For SNL-2 the imaging reveals a bright unresolved component to the source and this presents additional complexity due to possible AGN microlensing or variability. We tentatively use the relative magnification information to constrain the contribution from SNL-2's nearby companion galaxy, measuring a lensing mass of 12.59\,$\pm$\,0.30 $\times$\,10$^{10}$ M$_{\odot}$, a 9 per cent increase in mass. Our improved lens modelling reduces the mass uncertainty from 5 and 10 per cent to 2 and 3 per cent respectively. Our results support the conclusions of the previous analysis, with newly measured mass excess parameters of 1.17\,$\pm$\,0.09 and 0.96\,$\pm$\,0.10 for SNL-1 and SNL-2, relative to a Milky-Way like (Kroupa) initial mass function.


\end{abstract}





\section{Introduction}
\label{sec:Intro}

Although rare in the Universe, galaxy-scale strong gravitational lenses provide the most precise and accurate total mass measurements \citep{Treu2010,Courteau2014}. Unlike alternative mass estimators (such as stellar dynamics or spectroscopic analysis), strong lensing is unaffected by degeneracies due to the stellar population (e.g. age/metallicity), or structural properties (e.g. orbital anisotropy).

Gravitational lensing is only sensitive to the total (stars plus dark matter) mass projected within the Einstein radius (R$_{\rm Ein}$). The stellar component within early-type galaxies (ETGs) is generally more centrally concentrated than the dark matter (DM) halo, which follows an extended profile \citep[e.g.][]{NFW1996}. Therefore the fraction of mass contributed by the stars increases at smaller radius. Hence, for lenses in which the Einstein radius is small when compared to the galaxy's effective radius (R$_{\rm eff}$), lensing provides a robust measurement of the \textit{stellar} mass, which in turn provides information on the stellar initial mass function (IMF). In general this scenario is best realised with low redshift lenses, where the critical density for lensing is higher, and exceeded only at small physical radius.

The three lowest redshift (z\,$\lesssim$\,0.05) massive ETG lenses are ESO 325--G004, ESO 286--G022 and 2MASX J01414232--0735281. Whereas ESO 325--G004 was serendipitously discovered via HST imaging \citep{Smith2005,Smith2013}, the other two lenses were identified via targeted integral-field infrared spectroscopy with the SINFONI Nearby Elliptical Lens Locator Survey \citep[SNELLS,][hereafter SLC15]{SLC2015}; we refer to these three galaxy lens systems as SNL-0, SNL-1 and SNL-2 respectively. The \citetalias{SLC2015} lensing analysis of the SNELLS systems favoured a Milky-Way (MW) like (Kroupa) IMF \citep{Kroupa2001} and is strongly inconsistent with the `heavy' IMFs found in studies of massive ETGs via distant lenses, stellar dynamics, and direct spectral analysis \citep{Treu2010b,Cappellari2012,Conroy2012}.

The very local nature of the SNELLS sample allows the application of multiple independent IMF-determining techniques. High S/N optical spectra of the SNELLS lenses display features typical for a population of ETGs selected from SDSS to have similar velocity dispersions ($\sigma$ = 280 km s$^{-1}$) \citep[][figure 5]{Newman2017}. The spectral features for SNL-2 are found to be consistent with a MW-like IMF, in agreement with the lensing analysis, whereas SNL-1's spectra are found to favour a `heavy' IMF. Furthermore the stellar population synthesis modelling for SNL-1 finds a mass in excess of the total lensing mass estimated by \citetalias{SLC2015}. This result is further evidence of a tension between the results of different IMF estimators, motivating a more refined lens model for both galaxies.

The lensing mass of SNL-0 is robustly determined from earlier \textit{Hubble Space Telescope} (HST) observations \citep{Smith2013}. SNL-0's nearly complete Einstein ring lensing configuration provides accurate constraints on the mass model, with a 4 per cent uncertainty. By contrast SNL-1 and SNL-2 are two-image systems for which previous lensing analysis combined weak lensed-image positional constraints from SINFONI with low resolution 2MASS imaging \citepalias{SLC2015}. The estimated mass uncertainties were 5 and 10 per cent respectively due to the unknown contribution from external effects (i.e. shear).

In this paper we present improved mass estimates for SNL-1 and SNL-2, by exploiting recently acquired HST imaging. In Section \ref{sec:Data} we outline the data and our reduction procedures. In Section \ref{sec:LGS} we will provide a visual inspection of these galaxies, along with photometric analysis. In Section \ref{sec:LM} we analyse the lensing geometry with multiple parametric models, building upon the previous work and exploiting newly measured lensed image flux ratios. Finally in Section \ref{sec:discussion} we summarise these results and compare them to \citetalias{SLC2015}.

We use parameters from the 7-year \textit{Wilkinson Microwave Anisotropy Probe} (WMAP) when required, i.e. H$_0$ = 70.4 km s$^{-1}$ Mpc$^{-1}$, $\Omega _{\rm m}$ = 0.272 and $\Omega_{\Lambda}$ = 0.728 \citep{Komatsu2011}.

\section{Data}
\label{sec:Data}

\begin{table*}
	\centering
	\caption{Galaxy properties of SNL-1 and SNL-2. The magnitudes are quoted in the observed frame. The quoted luminosities are corrected for band-shifting, derived from \textsc{ezgal} \citep{Mancone2012}, and galactic extinction, from \citet{Schlafly2011}. }
	\label{tab:lens}
	\begin{tabular}{|c|c|c|l|}
		\hline
		Quantity & SNL-1 & SNL-2 & \multicolumn{1}{|c|}{Notes} \\
		\hline 
		NED ID & ESO 286-G022 & 2MASXJ01414232-0735281 &\\
		Lens z & 0.0312 & 0.0519 & \\
		$\sigma_{\rm 6dF}$  (km s$^{-1}$) &356\,$\pm$\,18& 320\,$\pm$\,18 & \citet{Campbell2014} \\
		$\sigma_{\rm e/2}$  (km s$^{-1}$) &289\,$\pm$\,14& 263\,$\pm$\,13 & \citet{Newman2017} \\
		Source z  & 0.926 & 1.969 & \vspace{2mm}  \\
		Fiducial aperture, R$_{\rm ap}$ (arcsec) & 2.38 & 2.21 & \citetalias{SLC2015}, SINFONI--based \\
		J($\leq$ R$_{\rm ap}$)  & 12.80 & 13.53  & PSF-corrected 2MASS \\				
		I$_{\rm F814W}$($\leq$ R$_{\rm ap}$)  & 13.85\,$\pm$\,0.02 & 14.53\,$\pm$\,0.02  &  \\ 
		L$_{\rm F814W}$($\leq$ R$_{\rm ap}$)  (10$^{10}$\,L$_{\odot}$) & 2.52\,$\pm$\,0.04 & 3.86\,$\pm$\,0.07  & extinction and band-shifting corrected \vspace{2mm} \\ 
		Half image-separation (arcsec) & 2.43\,$\pm$\,0.03 & 2.30\,$\pm$\,0.03 & this paper, HST--based\\
		I$_{\rm F814W}$($\leq$ separation)  & 13.83\,$\pm$\,0.02 & 14.50\,$\pm$\,0.02  & \vspace{2mm}  \\
		Flux Ratio (A/B)  & 2.2\,$\pm$\,0.1 & 2.5\,$\pm$\,0.1  &  \\
		\hline
		\end{tabular} 
\end{table*}

We observed SNL-1 and SNL-2 using HST Wide Field Camera 3 (WFC3), Uv-VISual (UVIS) channel, in GO cycle 23, (PI: Smith, R). We acquired three dithered F814W exposures for each target, for a total of 1050/1002\,sec respectively. We took a further three dithered exposures totalling 4413/4272\,sec, selecting a filter short of the 4000\,$\AA$ break for the lens-galaxy, but longer than any potential line-of-sight Ly\,$\alpha$ absorption in the source galaxy spectrum. Due to the differing redshifts of the lens-galaxies, SNL-1 was observed using F336W, and SNL-2 in F390W. 

We post-processed \textsc{calwf3} pipeline reduced UVIS data using the \textsc{astrodrizzle} software \citep{Gonzaga2012}. The images were drizzled onto a cosmic ray rejected final frame with a pixel scale of 0.025 arcsec/pix. Due to the limited number of frames in each passband, some artefacts remain after this process, which were masked in the subsequent analysis. 

\section{Lens and Source Properties}
\label{sec:LGS}

\begin{figure*}
	\centering
	\includegraphics[width=1.7\columnwidth]{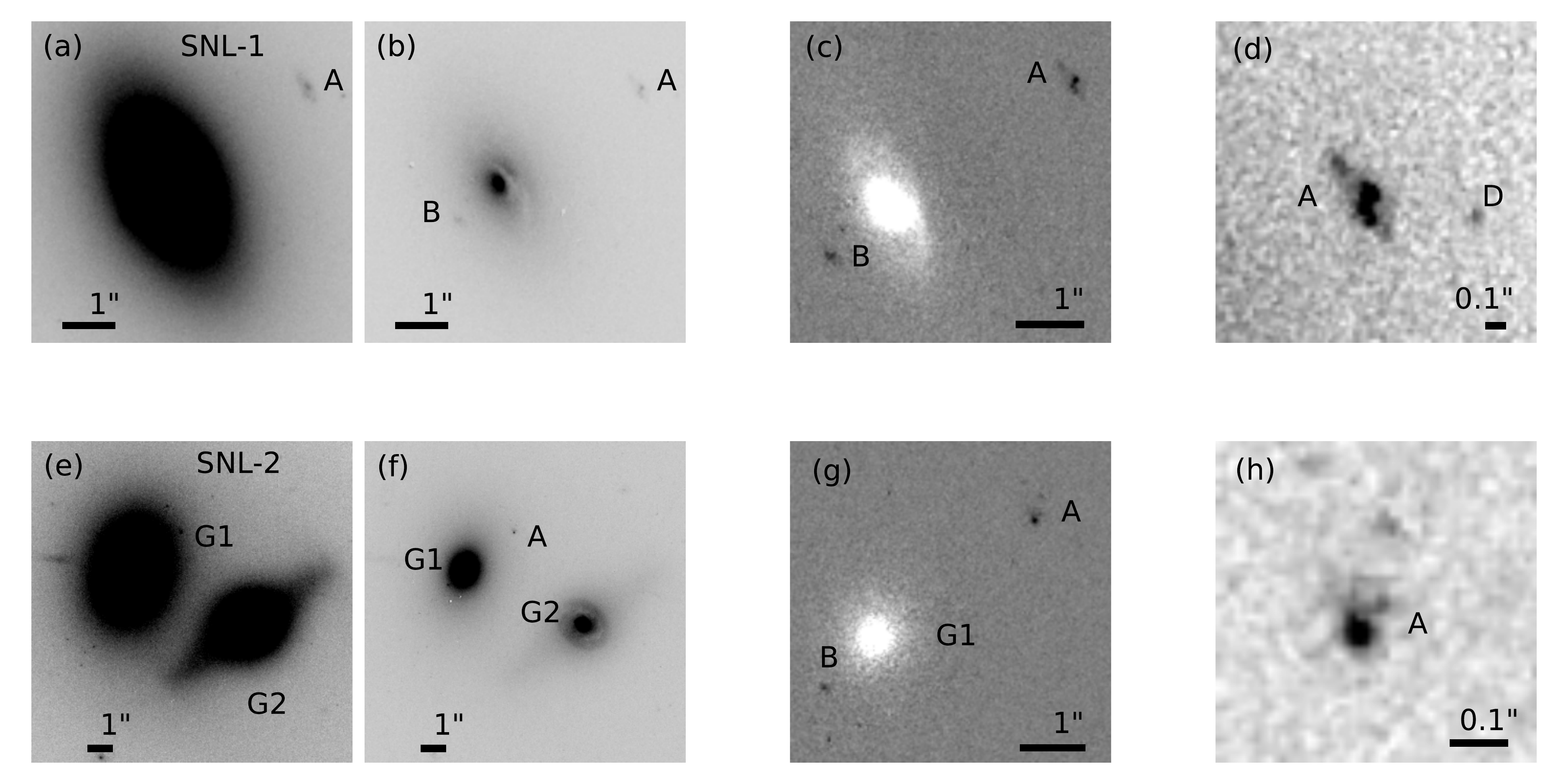}
	\caption{Panels (a-d) show SNL-1. A is the outer, and B the inner lensed image. (a) F814W image showing the lens ellipticity. (b) F336W image showing the inner obscuration from a disc. (c) scaled F336W with F814W subtracted for improved contrast of the lensed images. (d) detail of the outer image, A; there is a faint object D, which may or may not be associated. Panels (e-h) show SNL-2, G1 is SNL-2, G2 is the companion galaxy, with A the outer, and B the inner lensed image. (e) F814W image containing the lens and companion (f) F390W image showing the star formation ring in G2. (g) Scaled F390W with F814W subtracted for improved contrast of the lensed images. (h) the outer image, showing a compact core with a potentially associated diffuse structure.}
	
	\label{fig:panels}
\end{figure*}

In this section we report our measurements from the HST imaging. This includes: (a) morphological descriptions of the lens and the lensed sources, (b) improved measurements of the lensed-image positions and their relative fluxes, (c) independent measurements of the lens structural properties. The relevant parameters are summarised in Table \ref{tab:lens}.

\subsection*{SNL-1}

Extracted regions from the HST imaging for SNL-1 are shown in Figure \ref{fig:panels}(a-d). SNL-1 displays a regular E/S0 morphology, with an ellipticity of $\sim$\,0.4 (measured at the half-image separation from an \textsc{ellipse} fit), and slightly discy isophotes. SNL-1 was shown to be a fast rotator by \citet{Newman2017} and the HST imaging reveals dust obscuration within the central region ($\sim$\,1 arcsec, Figure \ref{fig:panels}b). This suggests the presence of a small cold ISM disc. 

The HST imaging confirms the two-image system discovered from previous SINFONI data. In Figure \ref{fig:panels}(d), we see internal structure in the background source. The outer image, A, displays a clumpy structure with a bright core, implying a late type galaxy. Little structure is visible in the inner image. From the new imaging we improve the locational constraints of the lensed images, and measure a half-image separation of 2.43\,$\pm$\,0.03 arcsec, which is 2 per cent larger than in \citetalias{SLC2015}. We derive the uncertainty from our ability to define centroids for the lensed images internal structure. The observed flux ratio (A/B) is 2.2$\pm$0.1, measured from aperture photometry within a lens galaxy subtracted image.

We measure the total lens flux with a two part model. The first component is a direct summation of the flux inside an elliptical aperture with a radius set at a preliminary estimate for R$_{\rm eff}$. The second component derives the flux contribution outside this region with a 1D S{\'e}rsic fit to the outer profile. We find the total magnitude to be I$_{\rm F814W}$ = 12.75\,$\pm$\,0.05, and a half-light radius of 3.90\,$\pm$\,0.03 arcsec. This is $\sim$\,20 per cent larger than the low S/N 2MASS-based R$_{\rm eff}$ measurement reported in \citetalias{SLC2015}.  

For consistency with previous work, we adopt a fiducial aperture of R$_{\rm ap}$\,=\,2.38 arcsec (the half-image separation derived by \citetalias{SLC2015}) when quoting magnitude measurements. We find I$_{\rm F814W}$($\leq$ R$_{\rm ap}$) = 13.85\,$\pm$\,0.02. Combined with the J band measured in \citetalias{SLC2015} we measure an (I$_{\rm F814W}$\,--\,J) colour of 1.05, which is consistent with the range of 1.01--1.07 derived for old metal-rich populations from synthesis models \citep{Conroy2009}.

\subsection*{SNL-2}

We present the HST imaging of SNL-2, in Figure \ref{fig:panels}(e-h). SNL-2 is confirmed to have an elliptical morphology, with a smooth light profile, and no discernible additional structure. However, SNL-2 lies with a nearby galaxy within a common, extensive and non-symmetric diffuse light halo. The companion, G2, located $\sim$\,7 arcsec away, is an edge on late-type galaxy, with disk and bulge components. A star forming ring within the companion's central bulge is seen in Figure \ref{fig:panels}(f). 

SNL-2 is confirmed to have a two-image lensing system, as found in the original SINFONI discovery data. The bright outer image, A, is a compact object, shown in Figure \ref{fig:panels}(h). We clearly observe a bright central region, with a tentatively associated low brightness structure extending outwards in a single direction. It is possible that an AGN dominates the flux of SNL-2's source. From the imaging, we measure a half image-separation of 2.30\,$\pm$\,0.03 arcsec (a 4 per cent increase on \citetalias{SLC2015}), with an uncertainty from our ability to centroid the unresolved source, and a flux ratio (A/B) of 2.5\,$\pm$\,0.1.

Due to the more complex local environment of SNL-2, we use \textsc{galfit} \citep{Peng2002} to model and subtract the companion. We model the companion with a S{\'e}rsic bulge, and a n\,$\approx$\,1 S{\'e}rsic disc. Then we fit SNL-2's light profile following the procedure used for SNL-1. We measure a total magnitude of I$_{\rm F814W}$ = 13.80\,$\pm$\,0.10, and an effective radius of 3.25\,$\pm$\,0.03 arcsec. This is significantly smaller than the 6 arcsec measured by \citetalias{SLC2015}, from low S/N 2MASS imaging. However, the complexity of SNL-2 with a bumpy, asymmetric light halo and companion galaxy limits the accuracy achievable when modelling this two-galaxy system. 

As with SNL-1, we measure the magnitudes within a fiducial radius, in this case adopted as R$_{\rm ap}$\,=\,2.21 arcsec, shown in Table \ref{tab:lens}. The (I$_{\rm F814W}$-J) colour of 1.00 is slightly bluer than the synthesis model predictions of 1.04--1.11 for old metal-rich populations \citep{Conroy2009}.

\section{Lens Modelling}
\label{sec:LM}

\begin{table}
	\centering
	\caption{Lens mass estimates from \textsc{lensmodel} in units of 10$^{10}$ M$_{\odot}$. Masses are measured within the fiducial radius, R$_{\rm ap}$. However the values are derived from the HST data and refined mass models.}
	\label{tab:lens_masses}
	
	\begin{tabular}{|c|c|c|c|}
		
		\hline 
		Model &  SNL-1 & \hspace{6mm}SNL-2 & Notes \\ 
		\hline 
		\multicolumn{4}{|c|}{Constrained by image positions, no shear} \\
		\hline
		\citetalias{SLC2015} & 9.27 & \hspace{6mm}13.07 &  \hspace{-3mm}table 2, No f$_{\rm{corr}}$ \\
		SIS  & 9.48 & \hspace{6mm}13.62 & \\ 
		SIE  & 8.78 & \hspace{6mm}14.11 &  \\
		MFL & 9.22 & \hspace{6mm}13.42 & \\ 
		\hline 
		\multicolumn{4}{|c|}{Constrained by flux ratio and positions, with shear (Shear)} \\
		\hline
		SIS + $\gamma$ & 9.08 & \hspace{6mm}-- & \hspace{-2mm}+4.3\,\% \\ 
		SIE + $\gamma$ & 9.57 & \hspace{6mm}-- & \hspace{-2mm}--8.8\,\% \\
		MFL + $\gamma$ & 9.41 & \hspace{6mm}-- & \hspace{-2mm}--2.1\,\% \\ 
		\hline 
		\multicolumn{4}{|c|}{Constrained by image positions with companion (Flux Ratio)} \\
		\hline
		MFL + SIS & -- & \hspace{6mm}12.83 & \hspace{-2mm}3.28  \\ 
		MFL + SIE & -- & \hspace{6mm}12.87 & \hspace{-2mm}3.26  \\
		MFL + MFL & -- & \hspace{6mm}13.15 & \hspace{-2mm}3.75 \\ 
		\hline 
		\multicolumn{4}{|c|}{Constrained by flux ratio, positions with companion} \\
		\hline
		MFL + SIS & -- & \hspace{6mm}12.28 &  \\ 
		MFL + SIE & -- & \hspace{6mm}12.38 &  \\
		MFL + MFL & -- & \hspace{6mm}12.69 & \\ 
		\hline
		\textbf{Adopted Mass} & \textbf{9.49$\pm$0.15} & \hspace{4mm}\textbf{12.59$\pm$0.30} & \hspace{-4mm} \\ 
		\hline
	\end{tabular} 
\end{table}

The main aim for this study, is to improve the mass estimates for SNL-1 and SNL-2 beyond the basic treatment in \citetalias{SLC2015}. We use the \textsc{lensmodel} code \citep{Keeton2001} to create parametrized profiles for each lensing system, informed by the HST imaging under the assumption that stellar mass dominates the lensing deflections. We measure the profiles' normalization, from which we extract the mass enclosed within the fiducial radius (R$_{\rm ap}$, from Table \ref{tab:lens}). In Section \ref{sec:PC} we apply only the improved image position constraints. Then in Section \ref{sec:FC} we include information from image-flux measurements to break the degeneracy between mass and external shear for SNL-1, and constrain the companion's effect in SNL-2. The results of our analysis are summarised in Table \ref{tab:lens_masses}.

\subsection{Positional constraints}
\label{sec:PC}

For SNL-1, we start our analysis with a singular isothermal sphere (SIS). This is the closest model to the spherical symmetry used in \citetalias{SLC2015}. From the SIS model we measure M($\leq$\,R$_{\rm ap}$)\,=\,9.48 $\times$\,10$^{10}$ M$_{\odot}$, 2 per cent larger than the previous estimate, due to the increased image separation. Figure \ref{fig:panels}(a) shows SNL-1 to have significant ellipticity, and to be orientated off axis to the image separation. We incorporate this using a singular isothermal ellipsoid (SIE) profile with a fixed ellipticity of 0.4. The resulting enclosed mass is $\sim$\,7 per cent smaller than predicted using a SIS model. 

As the stellar mass dominates within R$_{\rm Ein}$, we create a pixelised mass-follows-light (MFL) profile for the mass distribution. For the light profile we use an \textsc{iraf ellipse} fit \citep{Jed1987}, to minimise contamination from both the dust lane and the lensed images, as the basis for the mass profile. We assume the surface mass density is proportional to the surface brightness, such that the resulting model normalization is the mass-to-light ratio. The MFL estimated mass is $\sim$\,3 per cent smaller than the SIS model.

For SNL-2 we begin with a simplified case, in which we neglect the companion, and follow the procedure for SNL-1. The SIS model mass is $\sim$\,4 per cent larger than in \citetalias{SLC2015}, attributable to the larger HST measured image separation. We observe that SNL-2, like SNL-1, has a non-zero ellipticity and so model a SIE case, which increases the estimated mass by $\sim$\,4 per cent. We form a MFL profile for SNL-2, from an \textsc{ellipse} fit to a companion-subtracted F814W image, which will account for structure in the light profile. The MFL model estimates M($\leq$ R$_{\rm ap}$) to be $\sim$\,1.5 per cent smaller than SIS model. 

\subsection{Flux constraints}
\label{sec:FC}

\begin{figure}
	\centering
	\includegraphics[width=0.9\linewidth]{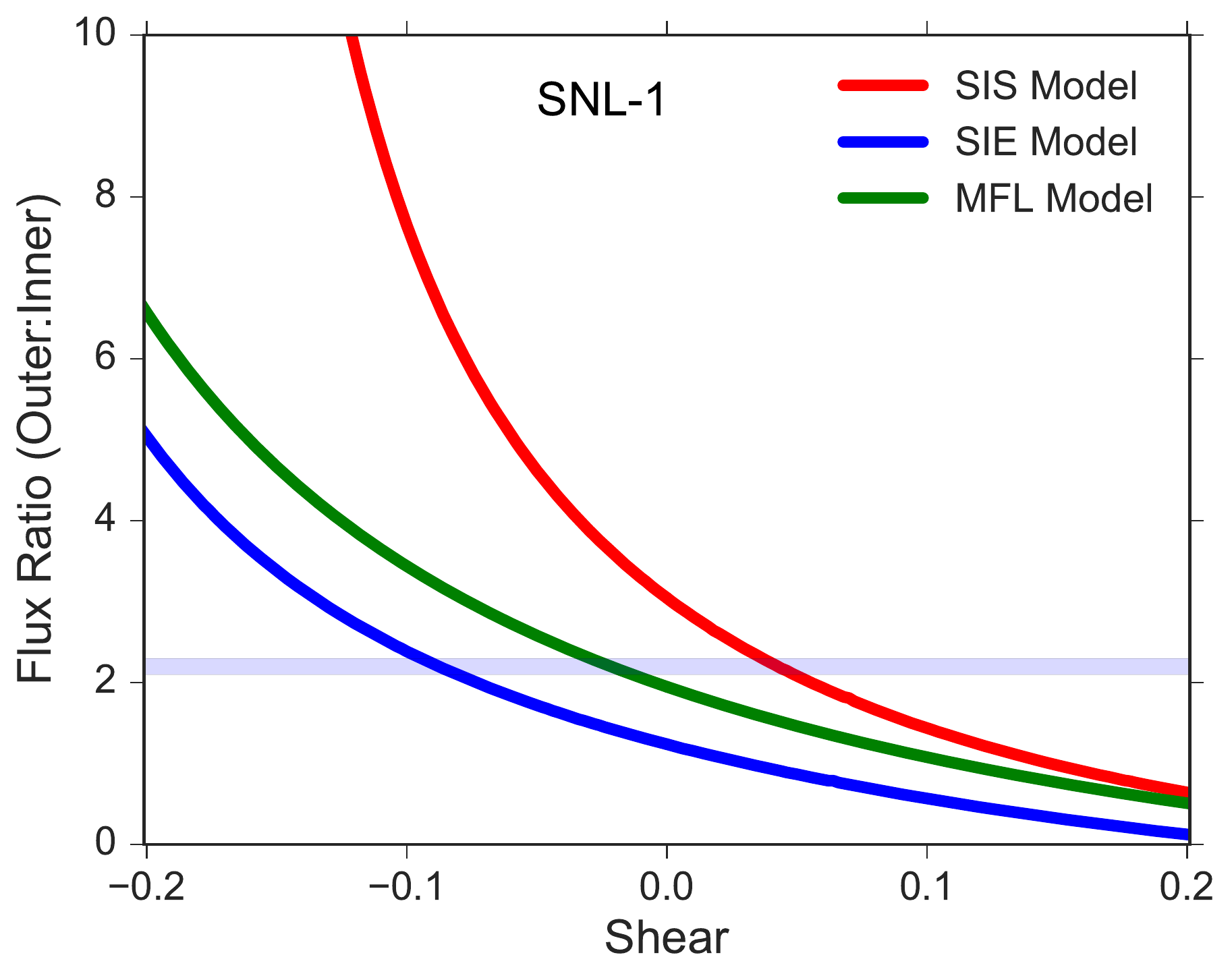}
	
	\caption{The external shear for given fixed image positions will modify the magnification each lensed image is subject to. We show this for SNL-1, comparing the predicted flux ratio of the lensed images to the shear. We shade in blue the measured flux ratio range of 2.2\,$\pm$\,0.1, indicating a weak compressive (negative) shear along the image-separation axis, for the SIE and MFL models.}
	\label{fig:snl1fratioshearmass}
\end{figure}

The previous lensing solutions assume an isolated lens galaxy. In reality the local environment causes a measurable effect on the lensing configuration, resulting in a degeneracy between mass and external shear ($\gamma$). An external shear causes an expansion along a given axis, and a perpendicular compression. We define a positive/negative shear to represent expansion/compression dominating in the image-separation axis, which reduces/increases the required enclosed mass for a given set of image positions. As the shear factor varies, the magnification of each image from a lensed source will change (illustrated for SNL-1 in Figure \ref{fig:snl1fratioshearmass}). The observed flux ratio, acting as a proxy for relative magnification of the lensed images, can therefore break this degeneracy between mass and shear. 

\begin{figure}
	\centering
	\includegraphics[width=0.9\linewidth]{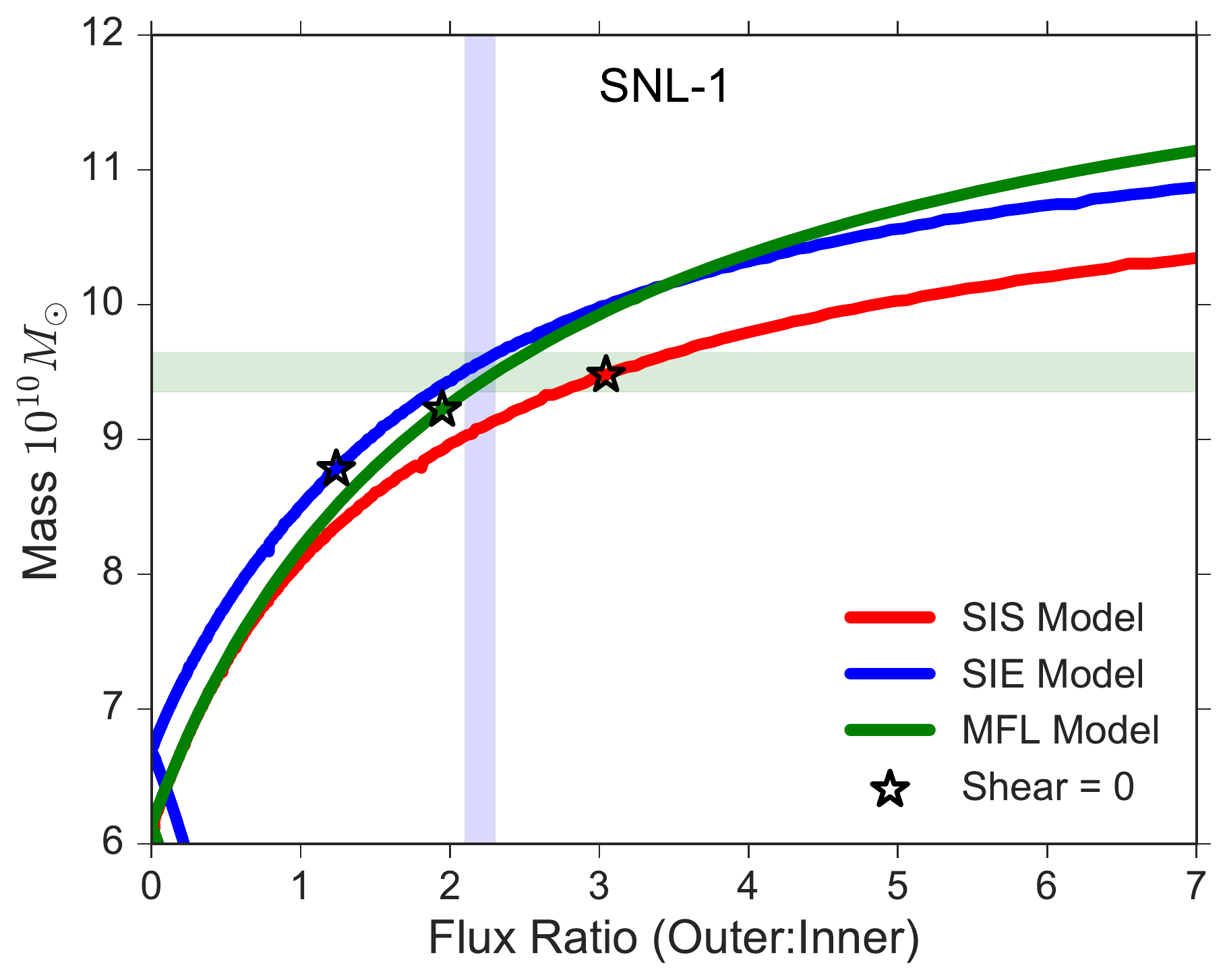}
	\caption{The model predicted flux ratio against mass for the three primary lens models of SNL-1. We shade the measured flux ratio of 2.20\,$\pm$\,0.1 in blue, from which we estimate the mass range breaking the mass-vs-shear degeneracy. The star symbols display the mass estimated for the case of no external shear. The mass range estimated by the SIE and MFL models for SNL-1 is shaded in green.}
	\label{fig:snl1massvfratiovspan}
\end{figure}

For SNL-1, we constrain the shear using the measured flux ratio (A/B) of 2.2\,$\pm$\,0.1, and show the resultant masses in Table \ref{tab:lens_masses}. For the SIE and MFL profiles this method recovers a compressive (negative) shear, increasing the measured lensing mass by $\sim$\,9 and $\sim$\,2 per cent respectively. (The SIS is less well defined, as the shear factor accounts for the combined effects of ellipticity, orientation and shear in this case.) Adopting the SIE+$\gamma$ and MFL+$\gamma$ models we derive limits on M($\leq$\,R$_{\rm ap}$) from model-to-model uncertainty, within the flux ratio bounds, to be 9.34\,--\,9.64 $\times$\,10$^{10}$ M$_{\odot}$, which is shaded in Figure \ref{fig:snl1massvfratiovspan}. 

\begin{figure}
	\centering
	\includegraphics[width=0.9\linewidth]{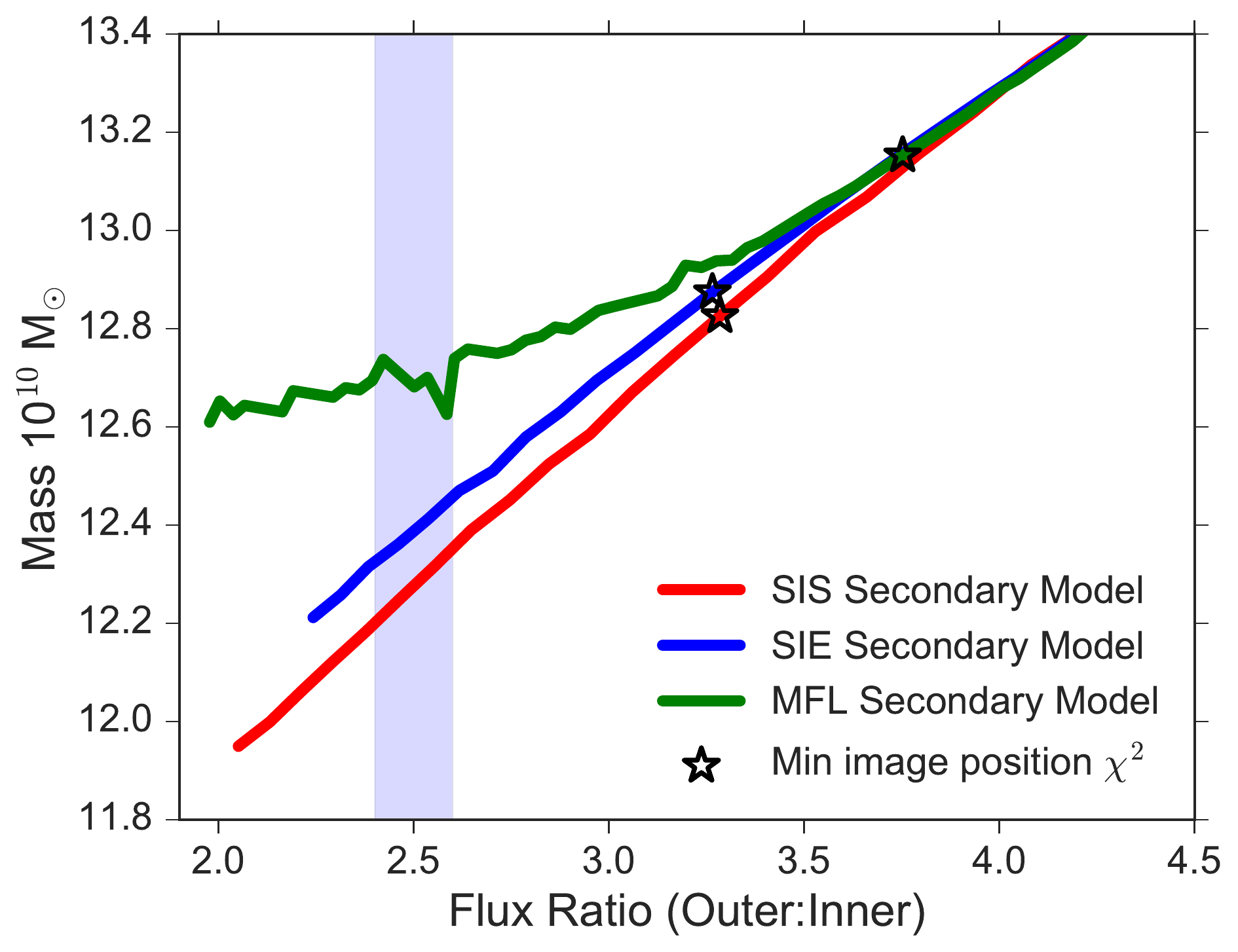}
	\caption{The modelled mass for SNL-2 against flux ratios for a varying mass companion. The shaded region defines the measured flux ratio of 2.5\,$\pm$\,0.1, with the black symbols indicating the best fit for the image positions. In order to fit the measured flux ratio, the offset in image position is $\sim$0.1 arcsec.}
	\label{fig:snl2_compnorm}
\end{figure}

For SNL-2, the close companion galaxy likely dominates the external mass distribution. We constrain the companion galaxy's contribution with the measured flux ratio, similar to the method for SNL-1. We treat SNL-2 with a MFL model throughout, and consider SIS, SIE and MFL descriptions for the secondary. We use these two-component models to predict the flux ratio as a function of the companion's mass normalization. In Figure \ref{fig:snl2_compnorm}, we compare the total lensing mass to the observed flux ratio (A/B), of 2.5\,$\pm$\,0.1. We find close agreement between the SIS and SIE secondary models, with significant divergence in the MFL case toward low flux ratios. As the companion is external to the primary lens configuration, we must consider its extended profile, and thus total (stars plus dark matter) mass. We therefore prefer the isothermal models for our treatment of the companion.

As shown in Figure \ref{fig:snl2_compnorm}, there are no models which match both the image positions and the observed image flux ratio. The models which reproduce the measured flux ratios yield a lensing mass of $\sim$\,12.33 $\times$\,10$^{10}$ M$_{\odot}$, but lead to an offset of $\sim$\,0.1 arcsec between the observed and predicted image positions. The models which best fit the image positions lead to a comparatively heavier primary lens, with a mass of $\sim$\,12.85 $\times$\,10$^{10}$ M$_{\odot}$. 

The deviation between measured and predicted flux ratios in our modelling, (2.5 and $\sim$\,3.3 respectively) seen in Figure \ref{fig:snl2_compnorm}, corresponds to $\sim$\,0.3 magnitudes. The compact nature of the source galaxy tentatively suggests a flux dominated by AGN activity, and for such a source microlensing can cause a discrepancy between the measured and predicted flux ratios of order a few tenths of a magnitude \citep[e.g.][]{Schechter2002, Schechter2014}. Alternatively a similar effect may result from intrinsic AGN variability combined with lensing path-length differences. These factors preclude obtaining an improved lensing mass estimate for this galaxy with the present data.

\section{Discussion and conclusion}
\label{sec:discussion}

The newly acquired HST data has revealed insights into the two-image lensing systems from SNELLS, including uncovering evidence for a previously unknown dust disc within the fast rotating SNL-1. We improved upon previous strong lensing analysis using the HST data to break the mass-vs-shear and companion degeneracies for SNL-1 and SNL-2 respectively. We measure precise lensed-image positions, and reliably quantify the lensed-image flux ratio, which was not possible with the SINFONI discovery data. We compare our adopted masses, shown in Table \ref{tab:lens_masses}, to those of \citetalias{SLC2015}, and estimate the stellar mass-to-light ratio ($\Upsilon$). Furthermore, combining the measured mass and luminosity in this paper, with the spectroscopically fit Kroupa reference stellar mass-to-light ratio ($\Upsilon_{\rm ref}$) \citep[][table 1]{Newman2017} (converted to I$_{\rm F814W}$ from the \textit{r}-band, with \textsc{ezgal}) we independently estimate the IMF mismatch parameter $\alpha$ within the fiducial radius (R$_{\rm ap}$). 

For SNL-1 we adopt a final lensing mass estimate of M($\leq$ R$_{\rm ap}$) = 9.49\,$\pm$\,0.15 $\times$\,10$^{10}$ M$_{\odot}$, from the SIE+$\gamma$ and MFL+$\gamma$ models. The quoted error is derived from the spread between lens profiles incorporating the flux ratio uncertainty. Due to the extended nature of the source, the contribution from microlensing is negligible. The positional errors provide a formal uncertainty of  $\leq$\,0.5 per cent. We obtain the stellar mass by subtracting the \textsc{eagle} DM mass contribution estimated in \citetalias{SLC2015} ($\sim$\,15 per cent) from the lensing mass. The estimated I$_{\rm F814W}$ measured $\Upsilon_{\rm R_{\rm ap}}$ is 3.21\,$\pm$\,0.12, and so combined with the converted $\Upsilon_{\rm ref}$ of 2.75 \citep{Newman2017}, we derive $\alpha$\,=\,1.17\,$\pm$\,0.09. This is 3 per cent smaller than 1.20\,$\pm$\,0.13 found by \citetalias{SLC2015}.

For SNL-2 the final lensing mass estimate is M($\leq$ R$_{\rm ap}$) = 12.59\,$\pm$\,0.30 $\times$\,10$^{10}$ M$_{\odot}$, derived from the MFL+SIS, and MFL+SIE models for the two regimes in Table \ref{tab:lens_masses}. The uncertainty is dominated by tension between the measured and predicted flux ratios. Following the \citetalias{SLC2015} \textsc{eagle} DM procedure, and incorporating the newly measured I$_{\rm F814W}$ luminosity we estimate $\Upsilon_{\rm R_{\rm ap}}$ to be 2.49\,$\pm$\,0.15. With the converted $\Upsilon_{\rm ref}$\,=\,2.59, we obtain $\alpha$\,$\simeq$ 0.96\,$\pm$\,0.10. This is a 2 per cent increase upon 0.94\,$\pm$\,0.17 measured by \citetalias{SLC2015}. 

In conclusion, our analysis of higher resolution and deeper imaging of SNL-1 and SNL-2 from HST supports the lensing masses, and the IMF $\alpha$ factors, estimated by \citetalias{SLC2015}. For SNL-2 further caution is required due to the complexity in modelling its source and companion galaxy. For SNL-1 the results show that the discrepancies in $\alpha$ reported by \citet{Newman2017} can not be attributed to the simplistic assumptions of the \citetalias{SLC2015} lens modelling. Future stellar- and gas-dynamical studies of SNL-1 should help to resolve this specific puzzle, and perhaps by implication begin to provide an explanation for the broader issue of agreement between the various methods for constraining the IMF.

\section*{Acknowledgements}

W. Collier was supported by an STFC studentship (ST/N50404X/1). RJS and JRL are supported by the STFC Durham Astronomy Consolidated Grant (ST/L00075X/1 and ST/P000541/1). Based on observations made with the NASA/ESA Hubble Space Telescope, obtained at the Space Telescope Science Institute, which is operated by the Association of Universities for Research in Astronomy, Inc., under NASA contract NAS 5-26555. These observations are associated with program 14210.






\bibliographystyle{mnras}
\bibliography{ref} 



%
%
%
%
%
\end{document}